\documentclass[nofootinbib,prl,superscriptaddress,twocolumn,showpacs]{revtex4}

\usepackage[utf8]{inputenc}
\usepackage{amssymb,amsthm,amsmath,amstext,amsbsy,bbm}
\usepackage{url}
\usepackage{nicefrac}
\usepackage{graphicx}
\usepackage{hyperref}

\newcommand{\cf}{\textit{cf.}}

\newcommand{\mathtext}[1]{\ \ \text{#1}\ \ }

\newcommand{\ket}[1]{\left|#1\right\rangle}

\newcommand{\ii}{\mathrm{i}}
\newcommand{\ee}{\mathrm{e}}

\newcommand{\RR}{\mathbb{R}}

\newcommand{\OO}{\mathcal{O}}
\newcommand{\YY}{Y}

\newcommand{\Laplace}{\Delta}

\newcommand{\vecr}{\mathbf{r}}

\newcommand{\vecn}{\mathbf{n}}

\newcommand{\vecq}{\mathbf{q}}

\newcommand{\vNabla}{\boldsymbol{\nabla}}

\newcommand{\del}[1]{\frac{\partial}{\partial #1}}

\newcommand{\ddr}{\mathrm{d}^3r}

\newcommand{\MeV}{\ensuremath{\mathrm{MeV}}}

\begin{document}

\title{Volume Dependence of Bound States with Angular Momentum}

\author{Sebastian König}
\affiliation{Helmholtz-Institut für Strahlen- und Kernphysik (Theorie)\\
and Bethe Center for Theoretical Physics, Universität Bonn, 53115 Bonn,
Germany\\[0.5em]}

\author{Dean Lee}
\affiliation{Department of Physics, North Carolina State University, Raleigh,
NC 27695, USA}
\affiliation{Helmholtz-Institut für Strahlen- und Kernphysik (Theorie)\\
and Bethe Center for Theoretical Physics, Universität Bonn, 53115 Bonn,
Germany\\[0.5em]}

\author{H.-W. Hammer}
\affiliation{Helmholtz-Institut für Strahlen- und Kernphysik (Theorie)\\
and Bethe Center for Theoretical Physics, Universität Bonn, 53115 Bonn,
Germany\\[0.5em]}

\date{\today}

\begin{abstract}
We derive general results for the mass shift of bound states with angular
momentum $\ell\geq1$ in a finite periodic volume.  Our results have direct 
applications to lattice simulations of hadronic molecules as well as atomic
nuclei.  While the binding of S-wave bound states increases at finite volume, we
show that the binding of P-wave bound states decreases.  The mass shift for
D-wave bound states as well as higher partial waves depends on the
representation of the cubic rotation group.  Nevertheless, the
multiplet-averaged mass shift for any angular momentum $\ell$ can be expressed
in a simple form, and the sign of the shift alternates for even and odd $\ell$.
We verify our analytical results with explicit numerical calculations.  We also
show numerically that similar volume corrections appear in three-body bound
states.
\end{abstract}

\maketitle

\paragraph{Introduction.}
With recent advances in computational power and algorithms, the physics of
numerous quantum few- and many-body systems can now be investigated from first
principles.  Lattice simulations are an important tool for such
calculations~\cite{Lee:2008fa,Bazavov:2009bb,Beane:2010em}.  The system is
solved numerically using a discretized spacetime in a finite volume.  In
practice, the finite volume is usually a cubic box with periodic boundaries.
The box modifies the bound state wave functions and leads to shifts in the
binding energies.  This shift needs to be subtracted from the calculated
energies for comparison to experiment.  In the case of S-wave bound states,
Lüscher has derived a formula for the finite volume mass shift of two-body
states~\cite{Luscher:1985dn}.  See Ref.~\cite{Beane:2010hg} for a recent
application of this method in lattice QCD to extract the mass of the proposed
H-dibaryon.

But there are also many bound states with nonzero orbital angular momentum.  In
nuclear physics, some particularly interesting examples occur in halo
nuclei~\cite{Riisager-94,Typel:2004zm,Rupak:2011nk,Hammer:2011ye}.  
These nuclei show a pronounced cluster structure. 
One-neutron halo nuclei can be regarded as a tightly-bound core with an extra
neutron.  In such cases the separation energy for the neutron is much smaller
than the binding energy of the core as well as the energy required for core
excitation.  Thus, the volume dependence of energy levels obtained in \textit{ab
initio} lattice calculations of such halo systems would behave as a two-body
system.

A well known example of a P-wave halo state is the $J^P=1/2^-$ excited state in
$^{11}$Be.  The electromagnetic properties of the low-lying states in $^{11}$Be
can be well described in a two-body halo 
picture~\cite{Typel:2004zm,Hammer:2011ye}.  If Coulomb interactions are
included, proton halos like $^8$B also become accessible.  In atomic physics,
several experiments have investigated strongly-interacting P-wave Feshbach
resonances in $^{6}$Li and
$^{40}$K~\cite{Regal:2003zz,Schunk:2005A,Gaebler:2007A} which can be tuned to
produce bound P-wave dimers.  There is interest in P-wave molecules in hadronic
physics \cite{Novikov:1977dq} as well as lattice investigations of the excited
nucleon spectrum in a number of different spin channels~\cite{Bulava:2009jb}.
Some of these states have been conjectured to have a molecular baryon--meson
structure~\cite{Matsuyama:2006rp}.  An extension of Lüscher's formula to higher
partial waves would provide a tool to discern molecular structures in hadronic
states as well as halo structures in nuclei from the finite-volume dependence of
lattice calculations for such systems.

In this letter, we derive general formulas for the finite-volume mass shift for 
bound states with nonzero orbital angular momentum $\ell$.  We also obtain a
simple expression for the multiplet-averaged mass shift for angular momentum
$\ell$.  We verify our analytical results with numerical calculations using an 
attractive short-range potential.  We note recent studies on the related topics
of extracting resonance properties and scattering phase shifts in higher partial
waves from finite-volume energy levels~\cite{Bernard:2008ax,Luu:2011ep}.  
Although our analytic derivation can be applied rigorously only to two-body
systems, we show numerically that quantitatively similar results also appear in
three-body systems.

\paragraph{Mass shift formula.}
In order to derive a general mass shift formula we consider a bound state
solution $\ket{\psi_B}$ of the Schrödinger equation
\begin{equation}
 \hat{H}\ket{\psi_B} = -{E_B}\ket{\psi_B} \mathtext{,}
 \hat{H} = -{\frac1{2\mu}}\Laplace_r+V(\vecr)
\label{eq:SG}
\end{equation}
with angular quantum numbers $(\ell,m)$ in a finite box of size $L^3$ with
periodic boundary conditions.  Following Lüscher's derivation
in~\cite{Luscher:1985dn}, the energy shift compared to the infinite volume
solution,
\begin{equation}
 \Delta m_B = E_B(\infty) - E_B(L) \,,
\label{eq:directdiff}
\end{equation}
can be written as
\begin{multline}
 \Delta m_B^{(\ell,m)} = \sum\limits_{|\vecn|=1}\int\ddr\,\psi_B^*(\vecr)\,
 V(\vecr)\,\psi_B(\vecr+\vecn L) \\ + \OO\big(\ee^{-{\sqrt2\kappa L}}\big) \,,
\label{eq:Delta_m-int-general}
\end{multline}
where $\vecn$ is an integer vector and $\kappa\equiv\sqrt{2\mu E_B}$ is the
binding momentum.  We assume that the potential has a finite range $R\ll L$.
Eq.~(\ref{eq:Delta_m-int-general}) arises from the overlap between copies of 
the system introduced by the periodic boundary conditions.  For $r>R$, the wave
function has the asymptotic form
\begin{equation}
 \psi_B(\vecr) =
 \YY_\ell^m(\theta,\phi)\,\frac{\ii^\ell\gamma\,\hat{h}_\ell^+(\ii\kappa r)}{r}
 \mathtext{,} \gamma\in\RR \,,
\label{eq:Asymptotic-WF}
\end{equation}
where $\hat{h}_\ell^+$ is a Riccati--Hankel function.  We will use the relation
\begin{equation}
 Y_\ell^m(\theta,\phi)\,\frac{\hat{h}_\ell^+(\ii\kappa r)}{r}
 = \left(-\ii\right)^\ell R_\ell^m\left(-\frac1\kappa\vNabla_r\right)
 \left[\frac{\ee^{-\kappa r}}{r}\right] \,,
\label{eq:SH-Hankel}
\end{equation}
which follows from Lemma B.1 in~\cite{Luscher:1990ux} and a derivative formula
for spherical Hankel functions~\cite{AbramStegPocket}.  The functions $R_\ell^m$
are the solid harmonics defined via
$R_\ell^m(\vecr) = r^\ell Y_\ell^m(\theta,\phi)$.\medskip

For S-waves, Eq.~(\ref{eq:SH-Hankel}) is a trivial identity.  Inserting the
Schrödinger equation to rewrite $V(\vecr)$ in \eqref{eq:Delta_m-int-general},
and using the fact that $\exp(-\kappa r)/(4\pi r)$ is a Green's function for the
operator $\left[\Laplace_r-\kappa^2\right]$, we recover Lüscher's result for
S-wave bound states.  In our notation this reads
\begin{equation}
 \Delta m_B = -{3}|\gamma|^2\,\frac{\ee^{-{\kappa L}}}{\mu L}
 + \OO\big(\ee^{-{\sqrt2\kappa L}}\big) \,.
\label{eq:Delta_m-S}
\end{equation}

We now generalize the mass shift formula to higher orbital angular momentum.  In
the following we explicitly show the derivation for P-waves.  We insert the
asymptotic expression \eqref{eq:Asymptotic-WF} with $\ell=1$ into
Eq.~(\ref{eq:Delta_m-int-general}) and use \eqref{eq:SH-Hankel} to rewrite the
Riccati--Hankel function.  For $m=0$ we find
\begin{multline}
 \Delta m_B^{(1,0)} = -\frac{\sqrt{3\pi}\gamma}{\mu\kappa}
 \sum\limits_{|\vecn|=1}\del{z}\psi_B^*(\vecr-\vecn L)\Big|_{\vecr=0} \\
 + \OO\big(\ee^{-{\sqrt2\kappa L}}\big)
\end{multline}
after integrating by parts.  For $m=\pm1$, the result is similar and involves
derivatives with respect to $x$ and $y$.  Evaluating the sums then yields
\begin{equation}
 \Delta m_B^{(1,0)} = \Delta m_B^{(1,\pm1)} = 3|\gamma|^2\,
 \frac{\ee^{-{\kappa L}}}{\mu L}
 + \OO\big(\ee^{-{\sqrt2\kappa L}}\big) \,.
\label{eq:Delta_m-P}
\end{equation}
Compared to the S-wave case, the P-wave mass shift is opposite in sign but equal
in magnitude.  Qualitatively, this means that S-wave bound states are more
deeply bound when put in a finite volume while P-wave states are less
bound.\medskip

For higher partial waves we can proceed in exactly the same manner.  The
results, however, are more complicated and the shift for D-waves and higher
partial waves depends on the quantum number $m$.  We note that due to the
periodic boundaries, the rotational symmetry group is reduced to a cubic
subgroup.  As a consequence, angular momentum multiplets are split into
irreducible representations of this subgroup (see, for example,
Ref.~\cite{Bernard:2008ax}).  A similar splitting also arises in lattice
calculations due to discretization artifacts.

The mass shift for general $\ell$ can be expressed as
\begin{equation}
 \Delta m_B = \alpha\left({\textstyle\frac{1}{\kappa L}}\right)
 \cdot|\gamma|^2\,\frac{\ee^{-{\kappa L}}}{\mu L}
 + \OO\big(\ee^{-{\sqrt2\kappa L}}\big) \,,
\label{eq:Delta_m-general}
\end{equation}
where the coefficients $\alpha\left(\frac{1}{\kappa L}\right)$ are given in 
Tab.~\ref{tab:Results} for $\ell=0,1,2$.  The irreducible representation of
the cubic group is denoted by $\Gamma$ in Tab.~\ref{tab:Results}.  A detailed
derivation of the general mass shift formula will be provided in a forthcoming
publication~\cite{Koenig:2011aa}.
\begin{table}[htbp]
 \begin{tabular}{c|c||c}
  $\ell$ & $\Gamma$ & $\alpha(x)$ \\
  \hline\hline
  $0$ & $A_1^+$ & $-3$ \\
  \hline
  $1$ & $T_1^-$ & $+3$ \\
  \hline
  $2$ & $T_2^+$ &
  $30x+135x^2+315x^3+315x^4$\\
  $2$ & $E^+$ &
  $\quad-\nicefrac{1}{2}\left(15+90x+405x^2+945x^3+945x^4\right)$\\
 \end{tabular}
\caption{Coefficient $\alpha(x)$ in the expression for the finite volume mass
shifts for $\ell=0,1,2$. $\Gamma$ indicates the corresponding representation of
the cubic group.}
\label{tab:Results}
\end{table}
The expressions for the finite volume mass shift become simpler when we sum
over all $m$ for a given $\ell$.  Using the trace formula for spherical
harmonics, it can be shown~\cite{Koenig:2011aa} that
\begin{multline}
 \sum_{m=-\ell}^\ell \Delta m_B^{(\ell,m)} =
 (-1)^{\ell+1}(2\ell+1)\cdot3|\gamma|^2\frac{\ee^{-\kappa L}}{\mu L} \\
 +\OO\big(\ee^{-{\sqrt2\kappa L}}\big) \,.
\label{eq:avshift}
\end{multline}
Dividing by $2\ell +1$, we obtain the average mass shift for states with angular
momentum $\ell$.  Apart from the overall sign, this average shift is independent
of $\ell$.  This follows from the fact that 
$Y_{\ell}^{m}(\theta,\phi)Y_{\ell}^{m\ast}(\theta,\phi)$ averaged over
$m=-\ell,\ldots,\ell$ is equal to $1/(4\pi)$ for all $\theta,\phi$, and $\ell.$
For the case $\ell=2$ (\cf~Tab.~\ref{tab:Results}), Eq.~\eqref{eq:avshift} can
be verified explicitly by averaging over the three-dimensional representation
$T_2^+$ and the two-dimensional representation $E^+$.  The mass shifts for the
S- and P-wave states are especially simple because these multiplets are not
split apart into more than one cubic representation.

\begin{figure}[htbp]
\centering
\includegraphics[width=0.35\textwidth,clip]{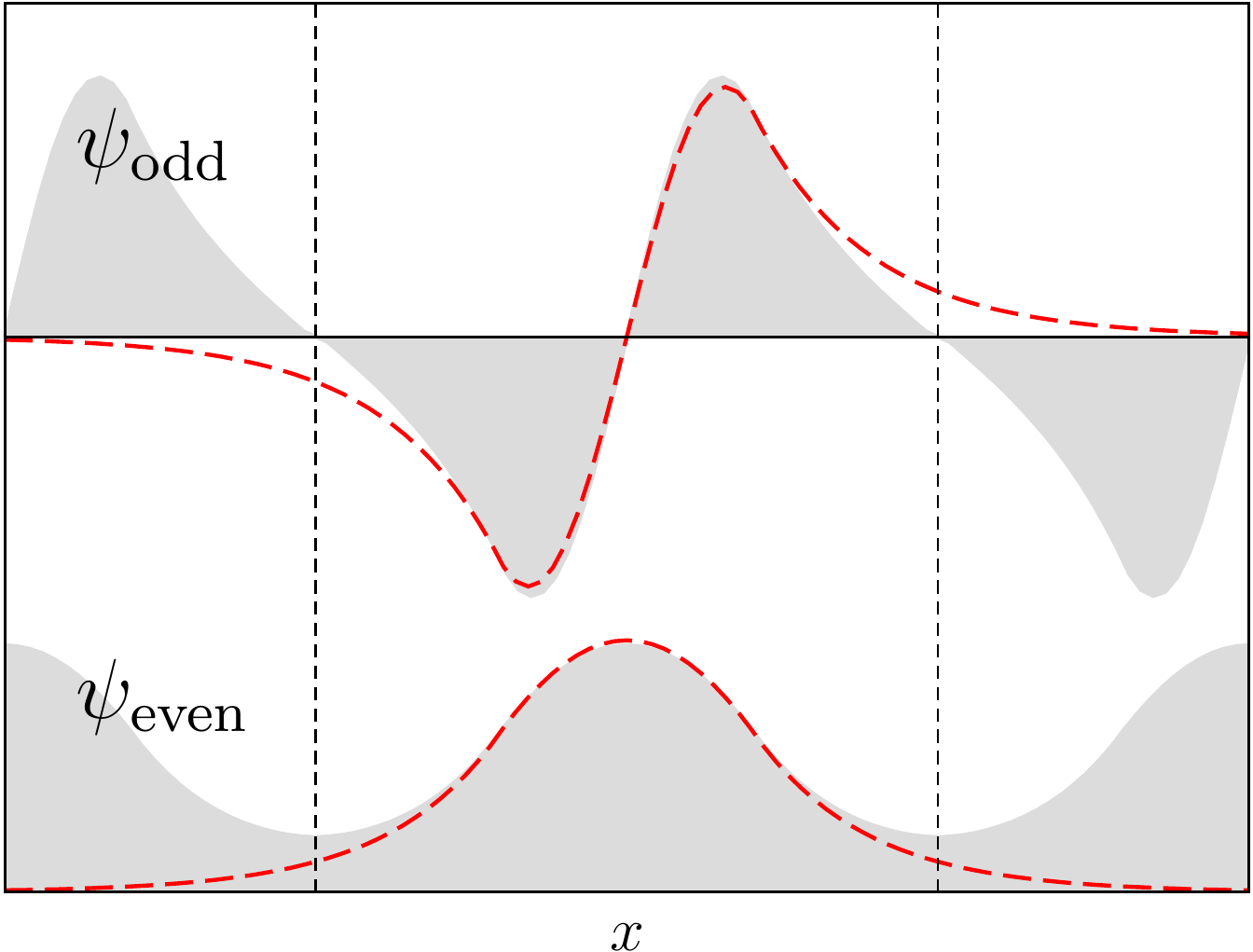}
\caption{(Color online) Wave functions with even (bottom) and odd parity (top)
for a one-dimensional square well potential in a box with periodic boundary
conditions.  The dashed lines give the infinite volume solutions for
comparison.}
\label{fig:WF}
\end{figure}

The sign of the finite volume mass shift can be explained in terms of the parity
of the wave function.  At infinite volume the tail of each bound state wave
function must vanish at infinity.  At finite volume, however, the bound state
wave functions with even parity along a given axis can remain nonzero
everywhere.  Only the derivative needs to vanish at the boundary, and the
kinetic energy is lowered by broadening of the wave function profile.  On the
other hand, a wave function with odd parity along a given axis must change sign
across the boundary.  In this case the wave function profile is compressed and
the kinetic energy is increased.  We have illustrated both cases for a
one-dimensional square well potential in Fig.~\ref{fig:WF}.

\begin{figure}[htbp]
\centering
\includegraphics[width=0.40\textwidth,clip]{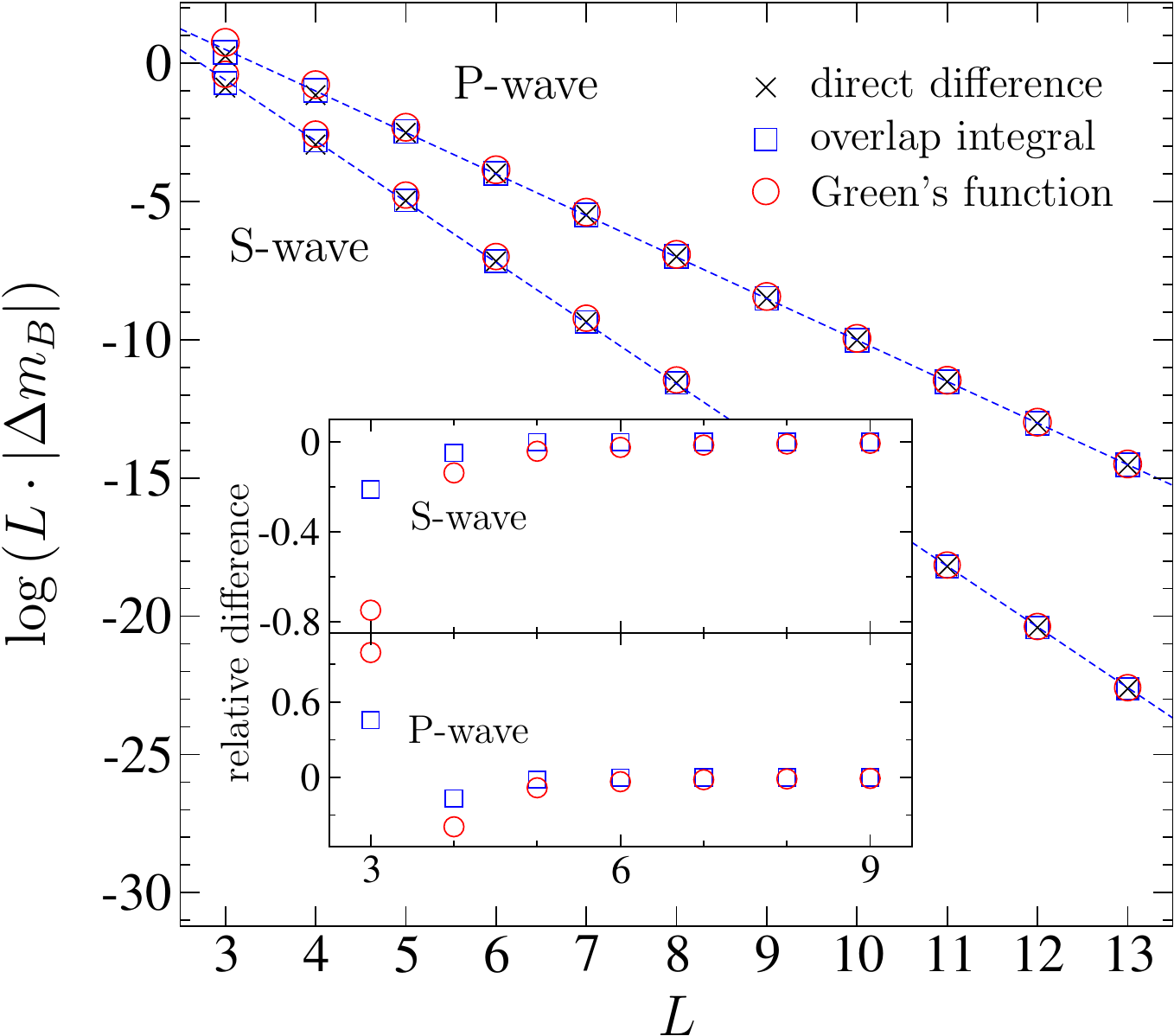}
\caption{(Color online) S- and P-wave mass shifts $\log(L\cdot|\Delta m_B|)$ as
functions of the box size $L$ (in lattice units).  We show the results obtained
from the direct difference Eq.~\eqref{eq:directdiff} (crosses), evaluation of
the overlap integral Eq.~\eqref{eq:Delta_m-int-general} (squares), and
discretized versions of Eqs.~\eqref{eq:Delta_m-S}, \eqref{eq:Delta_m-P}
(circles).  The dashed lines are linear fits to the squares.  In the inset, we
show the relative difference between the direct results and the overlap
(squares) and Green's function (circles) data.}
\label{fig:Shifts-with-rd}
\end{figure}

\paragraph{Comparison with numerical results.}
We test our predictions for the S-wave and P-wave mass shifts using numerical
lattice calculations.  In Fig.~\ref{fig:Shifts-with-rd}, we show the mass shifts
obtained from numerically solving the Schrödinger equation for a lattice
Gaussian potential
\begin{equation}
 V(r) = -V_0\,\exp\left(-r^2/(2R^2)\right)
\label{eq:V-Gauss}
\end{equation}
with $R=1$, $V_0 = 6$, and $\mu = 1$.  All quantities are in lattice units.
This potential does not have a finite range in a strict mathematical sense, but
the range corrections can be entirely neglected.  In order to compare the
dependence on the box size $L$ with the predicted behavior, we have plotted
$\log(L\cdot|\Delta m_B|)$ against $L$ (for S-waves $\Delta m_B$ is negative). 
The expected linear dependence is clearly visible.

For comparison we have calculated mass shifts using three different methods.
The crosses show the direct difference, Eq.~\eqref{eq:directdiff}, where we have
used $L_\infty=40$ to approximate the infinite volume solutions.  The boxes were
obtained from the overlap formula \eqref{eq:Delta_m-int-general}.  The circles
were calculated using discretized versions of \eqref{eq:Delta_m-S} and
\eqref{eq:Delta_m-P}, which we obtained by replacing $\exp(-\kappa r)/r$ with
the lattice Green's function
\begin{equation}
 G_\kappa(\vecn) = G\left(\vecn,-\frac{\kappa^2}{2\mu}\right)=\frac1{L^3}
 \sum\limits_{\vecq}\frac{\ee^{-\ii\vecq\cdot\vecn}}{Q^2(\vecq)+\kappa^2} \,,
\end{equation}
where $Q^2(\vecq) = 2\sum\nolimits_{i=1,2,3}(1-\cos q_i) \,$.  This
Green's function is also used to calculate the asymptotic normalization $\gamma$
from the lattice data.  This incorporates the correct dispersion
relation for our lattice model.  Both the overlap and Green's function results
were calculated using lattice wave functions from the $L_\infty=40$ calculation.

All three results agree well for both the S-wave and P-wave.  For small $L$
there are visible deviations which can be attributed to the
$\OO\big(\ee^{-{\sqrt2\kappa L}}\big)$-corrections as well as potential range
effects.  The inset in Fig.~\ref{fig:Shifts-with-rd} shows this more clearly. 
There we plot the relative differences between the (logarithmic) direct results
versus the overlap and Green's function data.

When we perform a linear fit to the overlap integral data (dashed lines in
Fig.~\ref{fig:Shifts-with-rd}) we obtain $\kappa = 2.198 \pm 0.005$, $|\gamma| =
11.5 \pm 0.2$ for the S-wave results; and $\kappa = 1.501 \pm 0.004$, $|\gamma|
= 7.0 \pm 0.1$ for the P-wave results.  The values for the asymptotic
normalization are in good agreement with the results $|\gamma| \sim 11.5$
(S-wave) and $|\gamma| \sim 7.2$ (P-wave) that are obtained directly from the
$L_\infty=40$ data.  Inserting the corresponding energy eigenvalues into the
lattice dispersion relation
\begin{equation}
 {-\mu E_B} = \left(1-\cos(-\ii\kappa)\right) \,,
\end{equation}
we find $\kappa \sim 2.211$ (S-wave) and $\kappa \sim 1.501$ (P-wave), again in
quite good agreement with the fit results.  The remaining small discrepancies
can be attributed to the mixing with higher partial waves induced by the lattice
discretization and the fact that we have not performed a continuum extrapolation
to vanishing lattice spacing.

\paragraph{Summary and outlook.}
In this letter, we have derived an explicit formula for the mass shift of P- and
D-wave bound states in a finite volume.  We have compared our results with
numerical calculations of the finite-volume dependence for a lattice Gaussian
potential and found good agreement with predictions.  For $\ell \geq 2$ the mass
shift depends on the angular momentum projection $m$ due to different
representations of the cubic group.  The average mass shift in a multiplet with
arbitrary angular momentum $\ell$ can be expressed in a simple way, and apart
from the alternating sign it is independent of $\ell$.  Applications to nuclear
halo systems such as $^{11}$Be and molecular states in atomic and hadronic
physics appear promising.  Our study provides a general framework for future
lattice studies of molecular states with angular momentum in systems with
short-range interactions.

Finite-volume dependence can be used to probe the structure of a number of 
nuclei with conjectured alpha-cluster substructures
\cite{Tohsaki:2001an,Chernykh:2007zz,Suzuki:2007wi}.  Recently, there have been
\textit{ab initio} lattice calculations of the low-lying states of $^{12}$C
using effective field theory \cite{Epelbaum:2009pd}.  In particular, the energy
for the spin-2 state of $^{12}$C was calculated and found in agreement with the
observed value of $-87.72~\MeV$, just a few MeV below the triple-alpha
threshold.  It is not known how angular momentum is distributed in this state,
and the study of finite volume effects may help to resolve this question.

Our results apply rigorously only to two-body systems.  However, there is
empirical evidence that Eq.~(\ref{eq:Delta_m-general}) also gives the asymptotic
$L$-dependence for three- and higher-body bound states at finite volume
\cite{Epelbaum:2009zsa,Kreuzer:2008bi}.  In these cases the values of $\kappa$
and $\gamma$ are fitted empirically. We can show this explicitly using an
extension of our Gaussian lattice model to three particle species.  We take the
particle masses to be equal, $m_1 = m_2 = m_3 = 2$, and consider Gaussian
two-body potentials of the form \eqref{eq:V-Gauss} with the same range, $R=2$,
but different interaction strengths $V_{0}^{12} = 2.5$, $V_{0}^{23} = 3.0$, and
$V_{0}^{31} = 3.5$ between the particles $12$, $23$, and $31$, respectively.

\begin{figure}[htbp]
\centering
\includegraphics[width=0.40\textwidth,clip]{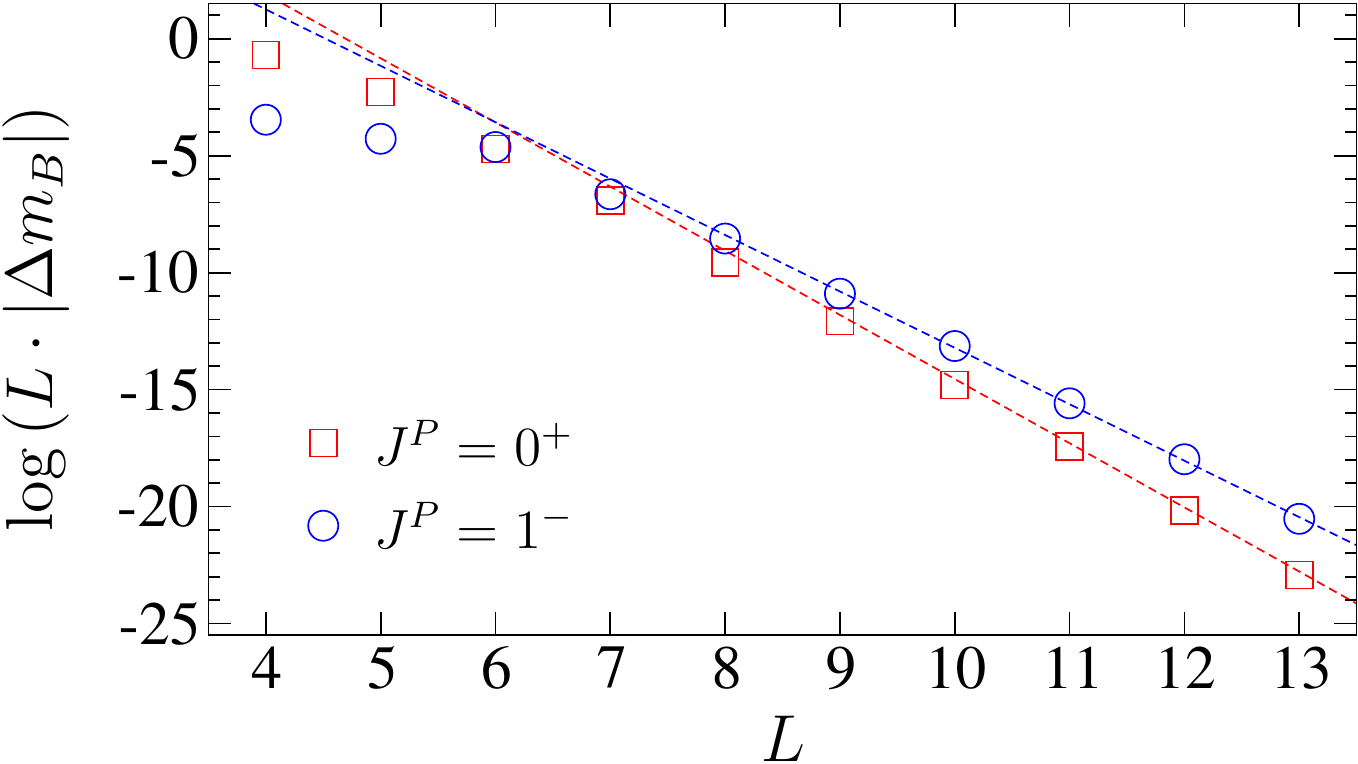}
\caption{(Color online) Trimer mass shifts $\log(-L\cdot|\Delta m_B|)$ as
functions of the box size $L$ (in lattice units).  We show lattice results for
the two lowest-lying trimer states with $J^P = 0^+$ (squares) and $J^P = 1^-$
(circles).}
\label{fig:Trimer}
\end{figure}

In Fig.~\ref{fig:Trimer}, we show the mass shifts for the lowest-lying 
trimer states with quantum numbers $J^P = 0^+$ and $J^P = 1^-$.  All
dimer states have less than half the binding energy of these trimers, which
indicates that there is no underlying two-body molecular structure.
%
%
As before we plot $\log(L\cdot|\Delta m_B|)$ against $L$.  Analogous to the
two-body case, we find that $\Delta m_B$ for $J^P = 0^+$ is negative, while
$\Delta m_B$ for $J^P = 1^-$ is positive.  We also find the same linear
dependence for $\log(L\cdot|\Delta m_B|)$ at large $L$.
Quite interesting are the subleading corrections which are especially strong for
$J^P = 1^-$ at smaller volumes.  This may be due to competing finite volume
corrections from negative S-wave and positive P-wave terms.  These results point
to a possible new application of finite volume corrections to probe the radial
distribution of angular momentum in complicated bound state systems.

\begin{acknowledgments}
This research was supported in part by the DFG through SFB/TR 16 ``Subnuclear
structure of matter'', the BMBF under contract No. 06BN9006, and by the US
Department of Energy under contract No. DE-FG02-03ER41260.
S.K. was supported by the ``Studien\-stiftung des deutschen Volkes'' and by the
Bonn-Cologne Graduate School of Physics and Astronomy.
\end{acknowledgments}

\end{document}